# VISCOPLASTIC PROPERTIES AND TRIBOLOGICAL BEHAVIOR OF DIAMOND-LIKE CARBON FILMS USING NANOINDENTATION AND NANOSCRATCH TESTS

V.Turq, J. Fontaine, J. L. Loubet, D. Mazuyer
*Ecole Centrale de Lyon, Laboratoire de Tribologie et Dynamique des Systèmes, UMR CNRS 5513*
*69134 Ecully Cedex, France*
*TEL : 33-472-186-287, Email: Viviane.Turq@ec-lyon.fr*

**Introduction**

Diamond-Like Carbon (DLC) films have been shown to demonstrate various tribological behaviors: in ultra-high vacuum (UHV), with either friction coefficients as low as 0.01 or less and very mild wear, or very high friction coefficients (>0.4) and drastic wear. These behaviors depend notably on gaseous environment, hydrogen content of the film [1], and on its viscoplastic properties [2,3]. A relation between superlow friction in UHV and viscoplasticity has indeed been established for a-C:H films and confirmed for a fluorinated sample (a-C:F:H). In this study, nanoindentation and nanoscratch tests were conducted in ambient air, using a nanoindentation apparatus, in order to evaluate tribological behaviors, as well as mechanical and viscoplastic properties of different amorphous carbon films.

**Experimental**

The samples were deposited on a Si (100) substrate by Plasma Enhanced Chemical Vapor Deposition (PECVD) process at different bias voltages, either from acetylene, cyclohexane precursors by d.c.-PECVD, or hexafluorobenzene mixed with hydrogen precursor by r.f.-PECVD for the fluorinated sample (a-C:F:H, noted FDLC). Details of the deposition process can be found in [4,5]. Thickness of the coating is 1µm, except for the FDLC, which is 0.4µm.

Nanoindentation and nanoscratch tests were carried out in ambient air, at room temperature, with a MTS NanoIndenter® XP apparatus. A spherical (radius 10µm) and a Berkovich diamond indenter were used. Mechanical and viscoplastic properties were evaluated from nanoindentation tests in continuous stiffness mode, using the Berkovich diamond indenter, with a maximum load of 100mN. As the load $P$ is applied exponentially as a function of time, the ratio between loading rate $P'$ and load $P$ is kept constant during indentation, and thus the strain rate $\dot{\varepsilon}$ is also constant. Five different ratios $P'/P$, from $3.10^{-3}$ up to $3.10^{-1}$ Hz, were used.
Nanotribological evaluation of the samples was conducted from nanoscratch tests at ramping load (0.1 to 10mN, 3 passes) and at constant load (5mN, 10 passes) with spherical diamond indenter.

**Results**

The strain rate sensitivity of the materials is estimated and fitted by a Norton-Hoff law: $H = H_0 \cdot \dot{\varepsilon}^x$ where $H$ is the hardness, $H_0$ a constant, $\varepsilon$ the strain and $x$ a constant called viscoplastic exponent (Table 1).
Contrary to UHV, no evidence of correlation between friction coefficients in ambient air and viscoplasticity can be made. But even in this environment, some very low friction coefficient values, as low as 0.04 (FDLC), with very mild wear have been evidenced (Figure 1). Hardness $H_0$ seems to be the key parameter: wear resistance in the air is improved with higher $H_0$ and friction coefficient decreases with $H_0$. Note that $H_0$ is also roughly linked with the hydrogen content of the coating for the non fluorinated samples, as it has been shown in [2]. The number of passes seems also to lead to a decrease of friction coefficient.

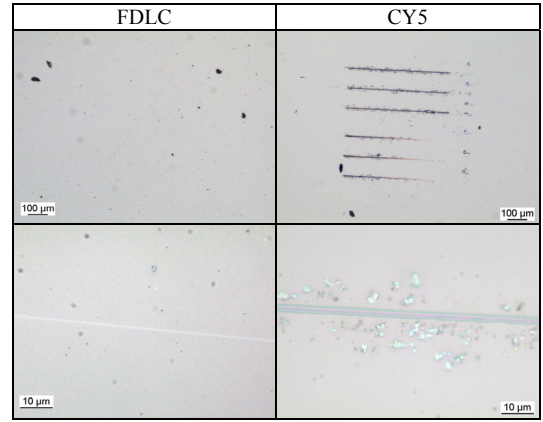

Figure 1: Constant load scratch micrograph

| Sample | H content (at. %) | $H_0$ (GPa) | x | µ ramping load | µ constant load | Wear |
|---|---|---|---|---|---|---|
| FDLC | 5/18(F) | 16 | 0.060 | 0.04-0.14 | 0.080 | ~ none |
| AC8 | 34 | 13 | 0.014 | 0.06-0.11 | 0.083 | ~ none |
| AC5 | 40 | 11 | 0.068 | 0.05-0.16 | 0.078 | mild |
| CY6.5 | 42 | 6.8 | 0.028 | 0.07-0.13 | 0.083 | mild |
| CY5 | 42 | 1.3 | 0.076 | 0.10-0.22 | 0.183 | severe |

Table 1: Summary of nanofriction tests results and viscoplastic properties

**Conclusion**

This study shows that in ambient air, wear resistance and frictional behavior of a-C:H and a-C:F:H samples is improved with hardness $H_0$. In UHV, the achievement of super-low friction is linked with the viscoplastic character. Thus, intermediary coating, with high hardness and viscoplastic exponent, as a-C:F:H will demonstrate satisfactory tribological behavior both in ambient air and in UHV.